\documentclass[page-classic,linenumbers]{epl2} 

\usepackage{amsmath}
\usepackage{multirow}

\title{Josephson effect in spin-singlet superconductor/ferromagnetic insulator/spin-triplet superconductor junctions with helical $p$-wave states}

\shorttitle{Josephson effect in spin-singlet superconductor/ferromagnetic insulator/spin-triplet}

\author{Q. Cheng\inst{1}\thanks{E-mail: \email{chengqiang07@mails.ucas.ac.cn}} \and B. Jin\inst{2}\thanks{E-mail: \email{biaojin@ucas.ac.cn}}}
\shortauthor{Q. Cheng \etal}

\institute{
  \inst{1} School of Science, Qingdao Technological University, Qingdao 266520, China\\
  \inst{2} School of Physics, University of Chinese Academy of Sciences, Beijing 100049, China
}

\abstract{
 We study the Josephson effect in spin-singlet superconductor/helical $p$-wave superconductor junctions with a ferromagnetic barrier using the quasiclassical Green function method. It is found that both $\sin{\phi}$-type and $\cos{\phi}$-type current-phase relations always exist, irrespective of the gap symmetries in superconductors. The indispensable condition for the $\sin{\phi}$-type and $\cos{\phi}$-type current is that the magnetization must have a component parallel to the crystallographic $a$ or $b$ axis, which is distinct from the case of $p$-wave superconductor described by a $\vect{d}$-vector with a uniform direction. The relation between the condition and the symmetries of the gap functions is analysed. We investigate in detail the symmetries and the sign reversal of the Josephson current when the magnetization is rotated.}
\begin{document}
\maketitle

\section{Introduction}
Since the discovery of the spin-triplet superconducting materials such as Sr$_{2}$RuO$_{4}$ \cite{Maeno, Mackenzie, Maeno2}, the Josephson effect in spin-singlet superconductor (SS)/spin-triplet superconductor (TS) and TS/TS junctions has been subjected to continuously growing interests \cite{Hasegawa, Barash, Sengupta, Lu, Brydon1, Brydon2, Bujnowski, Brydon3, Brydon4}. Generally, the Josephson current can be expressed as the composition of the harmonics $\sin{n\phi}$ and $\cos{n\phi}$, in which the number $n$ denotes the $n$th order contribution. In contrast to the Josephson junctions with the identical order parameter on the both sides \cite{Golubov}, SS/TS junctions do not permit the presence of the lowest order current (LOC) due to the orthogonality of the Cooper pair wave functions \cite{Pals, Millis, Asano}; the second-order term with the form of $\sin{2\phi}$ dominates the charge current-phase relation (CCPR). However, the situation will be changed when the interface is magnetically active. For example, the interface which is a ferromagnetic insulator (FI) or of spin-orbit coupling (SOC) can lead to the Josephson current in SS/TS junctions proportional to $\cos{\phi}$ when TS is in the chiral $p$-wave state \cite{Asano, Tanaka}. Moreover, it is found the reversal of the direction of the magnetization in FI will reverse the sign of the current. With the tunneling Hamiltonian perturbation theory, the authors in \cite{Brydon5} examine SS/FI/TS junctions in which the $\vect{d}$-vector with a uniform direction characterizes the $p$-wave states. The selection rules which permit LOC in the CCPRs and in the spin current-phase relations (SCPRs) are summarised. One of them is that the magnetization in FI must have a component parallel to the $\vect{d}$-vector of TS.

As the aforementioned $p$-wave states, the helical ones, $k_{x}\hat{x}\pm k_{y}\hat{y}$ and $k_{y}\hat{x}\pm k_{x}\hat{y}$, are also the candidate states for the spin-triplet superconducting material Sr$_{2}$RuO$_{4}$ \cite{Mackenzie, Maeno2, ZhangJ}. Among others, $k_{x}\hat{x}+k_{y}\hat{y}$ is the two-dimensional analog of the BW state (B phase) in ${}^{3}\text{H}_{\text{e}}$ \cite{Mackenzie, Maeno2}; $k_{x}\hat{y}-k_{y}\hat{x}$ is the favorable pairing state for the triplet part of the order parameter in the noncentrosymmetric superconductor CePt$_{3}$Si \cite{Settai,Bauer} which has been investigated from the topological viewpoint \cite{Sato, Tanaka2} and the aspect of spin current \cite{Chiken, Vorontsov}. One important difference between the aforementioned states and the helical ones is that the direction of $\vect{d}$-vector for the latter is not uniform and is pinned in $ab$ plane. The interplay of the helical superconductivity and ferromagnetism can bring novel transport properties. The charge conductance in ferromagnet/helical $p$-wave TS junctions have been clarified in \cite{Cheng}, which show strong anisotropic dependences and particular symmetries when one rotates the magnetization. The transport properties are very different from those in the junctions characterised by a $\vect{d}$-vector with a uniform direction \cite{Hirai, Zhang}. Thus, a novel Josephson effect is expected in SS/FI/helical $p$-wave TS junctions. What the selection rule for LOC is and how the current depends on the magnetization in FI are questions to be answered.

Motivated by these questions, we study in this paper the Josephson effect in SS/FI/helical $p$-wave TS junctions
using the quasiclassical Green function method with the Riccati parameterization \cite{Eschrig1, Eschrig2}. Both
$s$-wave and $d$-wave gap symmetries in SS are taken into account. We find the current possesses both sign reversal and symmetries when the magnetzation in FI is rotated; the condition that permits the LOC is the presence of the magnetization component parallel to $a$ or $b$-axis of the lattice in Sr$_{2}$RuO$_{4}$.
\section{Formalism}
\begin{figure}[h]
\begin{center}
\onefigure[height=6.5cm, width=8cm]{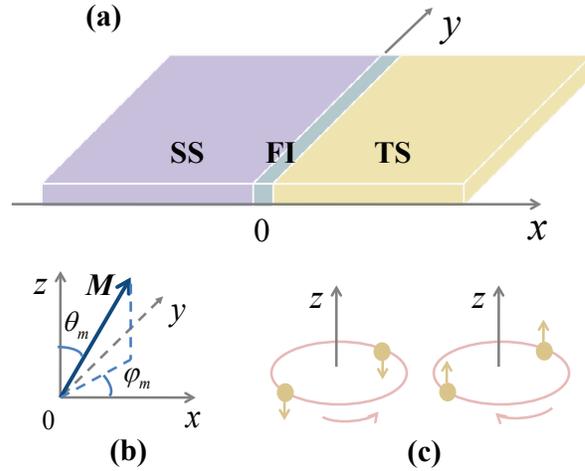}
\caption{(Colour online) (a): Schematic illustration of SS/FI/TS junctions. The $x (y)$-axis is defined by the crystallographic $a (b)$-axis. (b): The magnetization in FI specified by the polar angle $\theta_{m}$ and the azimuthal angle $\phi_{m}$. (c): The spins of Cooper pairs for the helical states which can be thought of as the superposition of the two states with spin parallel and anti-parallel to the $z (c)$-axis.}\label{fig.1}
\end{center}
\end{figure}

We consider the two-dimensional SS/FI/TS junctions which are assumed in clean limit as shown in fig.~\ref{fig.1}. The $\vect{d}$-vector in TS is taken as the following forms
\begin{equation}
\label{eq1}
\begin{split}
\vect{d}(\vect{k})=\Delta_{0}(k_{x},\pm k_{y},0) ~~~~~~\text{(HP}_{1\pm}-\text{wave}),\\
\vect{d}(\vect{k})=\Delta_{0}(k_{y},\pm k_{x},0) ~~~~~~\text{(HP}_{2\pm}-\text{wave}).
\end{split}
\end{equation}
The interface barrier, located at $x=0$ and along the $y$ axis, is modeled by a delta function $U(x)=(U_{0}+\vect{M}.\hat{\sigma})\delta(x)$ with the non-magnetic potential $U_{0}$ and the ferromagnetic term $\vect{M}\cdot\hat{\sigma}$. The magnetization $\vect{M}=M(\sin{\theta_{m}}\cos{\phi_{m}},\sin{\theta_{m}}\sin{\phi_{m}},\cos{\theta_{m}})$ with $\theta_{m}$ the polar angle and $\phi_{m}$ the azimuthal angle.

We first give the structure of the scattering matrix $\check{S}$ for the junctions in the normal state. The matrix is diagonal in the particle-hole space, i.e. $\check{S}=\mbox{diag}(\hat{S}, \hat{\tilde{S}})$
with
\begin{equation}
\label{eq3}
\hat{S}=\left(\begin{array}{cc}
S_{11}&S_{12}\\
S_{21}&S_{22}
\end{array}\right),~~~
\hat{\tilde{S}}=\left(\begin{array}{cc}
\tilde{S}_{11}&\tilde{S}_{12}\\
\tilde{S}_{21}&\tilde{S}_{22}
\end{array}\right).
\end{equation}
The $2\times2$ matrices $S_{11}(S_{22})$ and $S_{21}(S_{12})$ in spin space represent the electron reflection in the left (right) metal and the electron transition from the left (right) metal to the right (left) one, respectively. The hole reflection and transition are represented by the matrices $\tilde{S}_{11}(\tilde{S}_{22})$ and $\tilde{S}_{21}(\tilde{S}_{12})$. For the FI interface, we have $S_{22}=S_{11}, S_{12}=S_{21}=\hat{1}+S_{11}, \tilde{S}_{11}=\tilde{S}_{22}=S_{11}^*$ and $\tilde{S}_{12}=\tilde{S}_{21}=S_{12}^*$. The matrices can be expressed with the $x$ component of the wavevector of the incident electrons, the effective magnetization magnitude $X=\frac{Mm}{\hbar^2k_{F}}$ and the effective non-magnetic potential $Z=\frac{Um}{\hbar^2k_{F}}$.

The retard Green's function $\hat{g}$ in superconductors, which satisfies the Eilenberger equation \cite{Eilenberger}, can be written as
\begin{equation}
\label{eq5}
\hat{g}=-2\pi i\left(\begin{array}{cc}
g&f\\
-\tilde{f}&-\tilde{g}
\end{array}\right)+i\pi\hat{\tau_{3}},
\end{equation}
which is a $4\times4$ matrix in spin $\otimes$ particle-hole space. The quantities $g, f, \tilde{g}$ and $\tilde{f}$ can be parameterized as \cite{Eschrig2}
\begin{equation}
\label{eq6}
g=(1-\gamma\tilde{\gamma})^{-1},~~~f=(1-\gamma\tilde{\gamma})^{-1}\gamma,
~~~\tilde{g}=(1-\tilde{\gamma}\gamma)^{-1},~~~\tilde{f}=(1-\tilde{\gamma}\gamma)^{-1}
\tilde{\gamma},
\end{equation}
with the $2\times2$ coherence functions matrices $\gamma$ and $\tilde{\gamma}$. The equations for $\gamma$ and $\tilde{\gamma}$
are given by
\begin{equation}
\label{eq7}
\begin{split}
(i\hbar\vect{v}_{F}\cdot\nabla+2iw_{n})\gamma=\gamma\tilde{\Delta}\gamma-\Delta,\\
(i\hbar\vect{v}_{F}\cdot\nabla-2iw_{n})\tilde{\gamma}=\tilde{\gamma}\Delta\tilde{\gamma}-\tilde{\Delta},
\end{split}
\end{equation}
where $\vect{v}_{F}$ is the Fermi velocity, the Matsubara frenquency $w_{n}=2\pi k_{B}T(n+\frac{1}{2})$ and $\tilde{\Delta}(\vect{k})=[\Delta(-\vect{k})]^*$. For our model, the energy gap matrix in SS is $\Delta(\vect{k})=\Delta_{0}i\hat{\sigma_{y}}e^{i\phi}$ for $s$-wave symmetry and $\Delta(\vect{k})=\Delta_{0}(k_{x}^2-k_{y}^2)i\hat{\sigma_{y}}e^{i\phi}$ (or $\Delta_{0}k_{x}k_{y}i\hat{\sigma_{y}}e^{i\phi}$) for $d_{x^2-y^2}$ (or $d_{xy}$)-wave symmetry, respectively. The gap matrix in TS is $\Delta(\vect{k})=(\vect{d}(\vect{k})\cdot\hat{\sigma})i\hat{\sigma}_{y}$. The temperature dependence of the gap magnitude $\Delta_{0}$ is determined by the BCS-type equation with transition temperature $T_{C}$. $\phi$ gives the difference of the external phases of SS and TS when the phase in TS is taken as zero.

Ignoring the proximity effect, the incoming coherence functions $\gamma_{1(2)}$ and $\tilde{\gamma}_{1(2)}$ for SS (TS), bulk solutions in equilibrium, can be obtained through solving the eq.~(\ref{eq7}). The outgoing coherence function $\tilde{\Gamma}_{1}$ in SS can be given by the boundary condition \cite{Eschrig2}
\begin{equation}
\label{eq8}
\begin{split}
\tilde{\Gamma}_{1}=\tilde{\gamma}_{11}+\tilde{\gamma}_{12}(1-\gamma_{2}\tilde{\gamma}_{22})^{-1}\gamma_{2}\tilde{\gamma}_{21},
\end{split}
\end{equation}
with $\tilde{\gamma}_{\alpha,\beta}=\sum\limits_{\nu=1}^{2}\tilde{S}_{\alpha,\nu}\tilde{\gamma}_{\nu}S_{\nu,\beta}$, ($\alpha,\beta=1,2)$. The retard Green's function $\hat{g}$ in SS is obtained by substituting $\gamma(\tilde{\gamma})$ with $\gamma_{1}(\tilde{\Gamma}_{1})$ in eq.~(\ref{eq5}) and (\ref{eq6}).

The Josephson current density can be found from
\begin{equation}
\label{eq9}
\begin{split}
J&=-eN(0)k_{B}T\sum\limits_{w_{n}}\langle v_{Fx}j(w_{n},\theta)\rangle_{\text{FS}_{+}},\\
J_{S}&=\frac{\hbar}{2}N(0)k_{B}T\sum\limits_{w_{n}}\langle v_{Fx}j_S{}(w_{n},\theta)\rangle_{\text{FS}_{+}}
\end{split}
\end{equation}
where $j(w_{n},\theta)\equiv\mbox{Tr}[\hat{\tau}_{3}\hat{g}]$, $j_{S}(w_{n},\theta)\equiv\mbox{Tr}[\mbox{diag}(\sigma_{3},-\sigma_{3})\hat{g}]$, $N(0)$ is the density of states at the Fermi level in the normal state and the Fermi surface average is only
over positive directions. The dimensionless charge (spin) current denoted by $I_{J}(I_{S})$ can be expressed as $I_{J}=\frac{eIR_{N}}{k_{B}T_{C}}(I_{S}=\frac{e^2I_{S}R_{N}}{\hbar k_{B}T_{C}})$, where $I(_{S})=J(_{S})A$ is the current for junctions with interface area $A$ and $R_{N}$ is the resistance for junctions in the normal state.
\section{Results and discussions}
First, we present the results for the junctions without the non-magnetic potential, i.e., $Z=0$. In this case,
we have the general consequences of the charge current which are
\begin{align}
\label{eq10}
I_{J}(\theta_{m},\phi_{m}, \phi)=I_{J}(\pi-\theta_{m}, \phi_{m},\phi),\\
\label{eq11}
I_{J}(\theta_{m},\phi_{m},\phi)=I_{J}(\theta_{m},\pi+\phi_{m},\pi+\phi),\\
\label{eq12}
I_{J}(\theta_{m},\phi_{m},\phi)=-I_{J}(\pi-\theta_{m},\pi+\phi_{m},2\pi-\phi).
\end{align}
Eq.~(\ref{eq10}) shows the symmetry of the current on the reflection of the magnetization about the $xy$-plane; eq.~(\ref{eq11}) implies a symmetry of the current when $\vect{M}$ is rotated by 180 degrees with respect to the $z$-axis and the phase is simultaneously shifted by $\pi$; eq.~(\ref{eq12}) means the reversal of the magnetization in FI will reverse the sign of the current which is also found in SS/FI/TS junctions with TS described by a $\vect{d}$-vector with a uniform direction \cite{Tanaka, Brydon5}. The three consequences are common properties of the charge current for SS/FI/TS junctions with helical $p$-wave states. The symmetry and sign reversal of the spin current $I_{S}$ will be discussed later.

\begin{figure}[h]
\begin{center}
\onefigure[height=7cm, width=8cm]{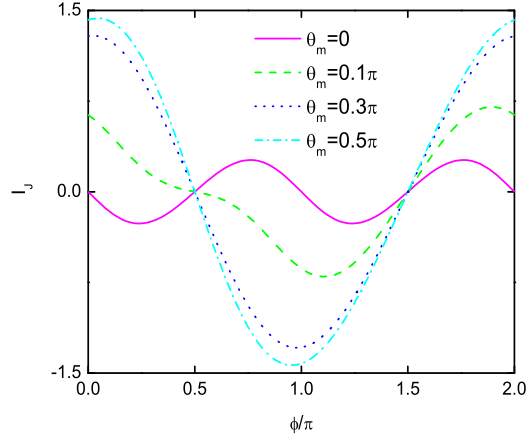}
\caption{(Colour online) The CCPRs for $\phi_{m}=0$, $Z=0$, $X=1$ and $T=0.3T_{C}$ in the $s$-wave SS/HP$_{1\pm}$-wave TS junctions.}\label{fig.2}
\end{center}
\end{figure}

\begin{figure}[h]
\begin{center}
\onefigure[height=4cm, width=8cm]{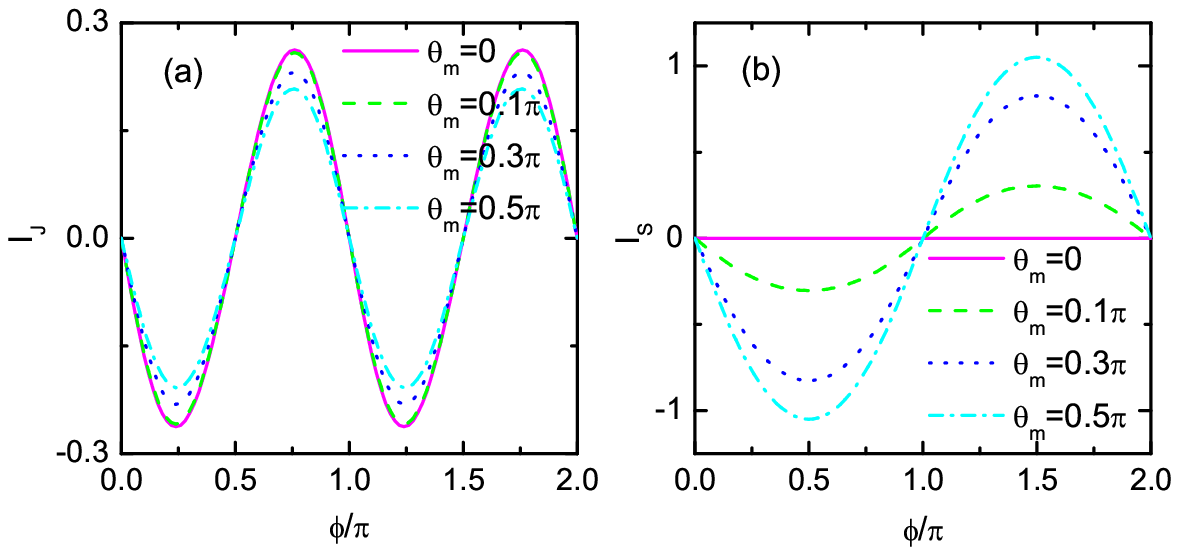}
\caption{(Colour online) (a): The CCPRs for $\phi_{m}=0.5\pi$, $Z=0$, $X=1$ and $T=0.3T_{C}$ in the $s$-wave SS/ HP$_{1\pm}$-wave TS junctions. (b) The corresponding SCPRs with the same parameters.}\label{fig.3}
\end{center}
\end{figure}

Now, we give the numerical results for the $s$-wave SS/HP$_{1\pm}$-wave TS junctions which possess the same Josephson current. Fig.~\ref{fig.2} shows the CCPRs for $\phi_{m}=0$ with various $\theta_{m}$ at $T=0.3T_{C}$. When $\theta_{m}=0$, i.e., $\vect{M}\perp\ \vect{d}$, the current is proportional to $\sin{2\phi}$, a typical relation for SS/TS junctions. Once $\theta_{m}$ deviates from $0$, the lowest order contribution with the $2\pi$-period starts to emerge. As $\theta_{m}$ tends to $0.5\pi$, the CCPR is dominated by $\cos{\phi}$. Since the spin current is zero for $\phi_{m}=0$, we do not show the SCPRs in fig.~\ref{fig.2}. Plotted in fig.~\ref{fig.3}(a) and (b) are the CCPRs and the SCPRs, respectively, for $\phi_{m}=0.5\pi$. Different from the case of $\phi_{m}=0$, the CCPRs for all $\theta_{m}$ show themselves as $\sin{2\phi}$ line shapes which indicate the absence of LOC when the magnetization is in the $yz$-plane with no $x$-component. The spin current in this situation is dominated by the harmonic $\sin{\phi}$ when $\theta_{m}\neq0$.

\begin{figure}[h]
\begin{center}
\onefigure[height=4cm, width=8cm]{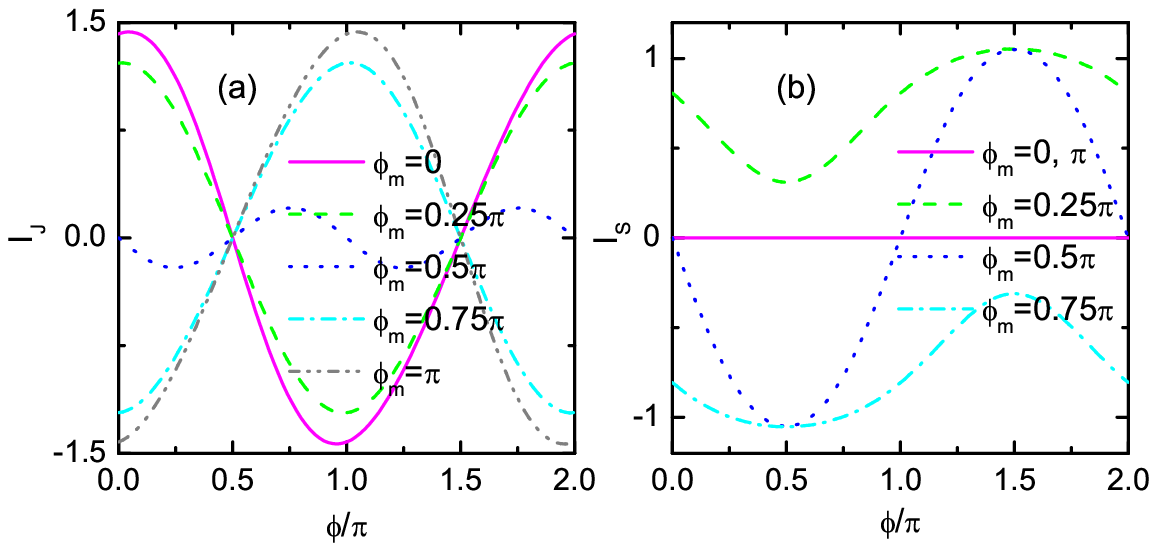}
\caption{(Colour online) (a): The CCPRs for $\theta_{m}=0.5\pi$, $Z=0$, $X=1$ and $T=0.3T_{C}$ in the $s$-wave SS /HP$_{1\pm}$-wave TS junctions. (b): The corresponding SCPRs with the same parameters.}\label{fig.4}
\end{center}
\end{figure}

Fig.~\ref{fig.4} shows the CCPRs and the SCPRs for $\theta_{m}=0.5\pi$, i.e. $\vect{M}$ in the $xy$-plane, with various $\phi_{m}$. The $\cos{\phi}$-type LOC in the charge current will appear once $\vect{M}$ deviates from the $y$-axis. When the magnetization sweeps across the $y$-axis, the sign of the charge current will be reversed. Actually, we have the equality
\begin{equation}
\label{eq13}
I_{J}(\theta_{m},\phi_{m},\phi)=-I_{J}(\theta_{m},\pi-\phi_{m},2\pi-\phi),
\end{equation}
for any $\theta_{m}$. Magnetizations with opposite $x$ component will create charge currents with opposite directions. We do not plot the CCPRs for $\pi<\phi_{m}<2\pi$ because we also have the equality
\begin{equation}
\label{eq14}
I_{J}(\theta_{m}, \phi_{m}, \phi)=I_{J}(\theta_{m}, 2\pi-\phi_{m}, \phi).
\end{equation}
The equalities in eq.~(\ref{eq13}) and (\ref{eq14}) are extra sign reversal and symmetry satisfied by $I_{J}$ for the $s$-wave SS/HP$_{1\pm}$-wave TS junctions besides the general consequences presented in eq.~(\ref{eq10})-(\ref{eq12}). The spin current $I_{S}$ still keep the $\sin{\phi}$ form for $\theta_{m}\neq0$ as shown in fig.~\ref{fig.4}(b).

From the CCPRs presented in fig.~\ref{fig.2}-\ref{fig.4}, we can summarise the condition for the presence of the $\cos{\phi}$-type LOC in the charge current which is that the magnetization must have the $x$-component, i.e. $\vect{M}\cdot\hat{x}\neq0$. The condition for the presence of the $\sin{\phi}$-type LOC in the spin current is $\vect{M}\cdot\hat{y}\neq0$ which can be summarised in a similar way.

For the s-wave SS/HP$_{2\pm}$-wave TS junction, the condition for LOC in the charge current will turn into the indispensable presence of the $y$ component of the magnetization, i.e. $\vect{M}\cdot\hat{y}\neq0$. Magnetizations with opposite $y$ component will bring charge currents with opposite directions,
\begin{equation}
\label{eq15}
I_{J}(\theta_{m}, \phi_{m},\phi)=-I_{J}(\theta_{m},2\pi-\phi_{m}, 2\pi-\phi),
\end{equation}
for any $\theta_{m}$. The extra symmetry will become
\begin{equation}
\label{eq16}
I_{J}(\theta_{m}, \phi_{m}, \phi)=I_{J}(\theta_{m}, \pi-\phi_{m},\phi).
\end{equation}
The condition for the LOC in the spin current of the junction turns into $\vect{M}\cdot\hat{x}\neq0$.
These differences between the CCPRs (or SCPRs) for HP$_{1\pm}$-wave and those for HP$_{2\pm}$-wave are a result of the $\frac{\pi}{2}$ difference between the $x(y)$ component and the $y(x)$ component.

Although the selection rules for the $\cos{\phi}$-type LOC in CCPRs and for the $\sin{\phi}$-type LOC in SCPRs are obtained from the current of junctions without the non-magnetic potential, they also apply to the junctions with $Z\neq0$. According to the rules, we can divide the junctions into two groups which are given in table. 1. For $Z=0$, the CCPRs of junctions in the group with the rule $\vect{M}\cdot\hat{x}\neq0$ obey the sign reversal and symmetry presented in eq.~(\ref{eq13}) and (\ref{eq14}); the CCPRs of junctions in the other group obey the relations presented in eq.~(\ref{eq15}) and (\ref{eq16}). In contrast, the Josephson current calculated in \cite{Brydon5} is only dependent on the relative angle of the $\vect{d}$-vector and the magnetization. That means the current is independent of the azimuthal angle of the magnetization when $\vect{d}\parallel\hat{z}$. In order to obtain the $\cos{\phi}$-type LOC there, two conditions must be satisfied: the magnetization component parallel to the $\vect{d}$-vector and the same parity of the gaps with respect to the interface momentum. As a result, the emergence of the $\cos{\phi}(\sin{\phi})$-type LOC in the charge (spin) current is restricted to the junctions with $s$ or $d_{x^2-y^2}$ ($d_{xy}$) symmetry in SS and $p_{x}$ or $p_{x}+ip_{y}$ ($p_{y}$) symmetry in TS.
\begin{table}
\caption{The selection rules for $\sin{\phi}$-type and $\cos{\phi}$-type LOC in SS/FI/Helica $p$-wave TS junctions.}
\label{table.1}
\begin{center}
\renewcommand{\multirowsetup}{}
\begin{tabular}{clccc}
\hline\hline
Current & SS & TS & $\cos{\phi}$ & $\sin{\phi}$ \\
\hline
\multirow{4}{1cm}{I$_J$} & $s$, $d_{x^2-y^2}$ & $k_{x}\hat{x}\pm k_{y}\hat{y}$ & \multirow{2}{2cm}{$\vect{M}\cdot\hat{x}\neq0$} & \multirow{2}{5cm}{$Z\neq0,\vect{M}\cdot\hat{y}\neq0$ and $\vect{M}\cdot\hat{z}\neq0$} \\
 & $d_{xy}$ & $k_{y}\hat{x}\pm k_{x}\hat{y}$ & &  \\
 & $s$, $d_{x^2-y^2}$ & $k_{y}\hat{x}\pm k_{x}\hat{y}$ & \multirow{2}{2cm}{$\vect{M}\cdot\hat{y}\neq0$} & \multirow{2}{5cm}{$Z\neq0,\vect{M}\cdot\hat{x}\neq0$ and $\vect{M}\cdot\hat{z}\neq0$} \\
 & $d_{xy}$ & $k_{x}\hat{x}\pm k_{y}\hat{y}$ &  & \\
\hline
 Current & SS & TS & $\sin{\phi}$ & $\cos{\phi}$ \\
\hline
\multirow{4}{1cm}{I$_S$} & $s$, $d_{x^2-y^2}$ & $k_{x}\hat{x}\pm k_{y}\hat{y}$ & \multirow{2}{2cm}{$\vect{M}\cdot\hat{y}\neq0$} & \multirow{2}{5cm}{$Z\neq0,\vect{M}\cdot\hat{x}\neq0$ and $\vect{M}\cdot\hat{z}\neq0$} \\
 & $d_{xy}$ & $k_{y}\hat{x}\pm k_{x}\hat{y}$ & &  \\
 & $s$, $d_{x^2-y^2}$ & $k_{y}\hat{x}\pm k_{x}\hat{y}$ & \multirow{2}{2cm}{$\vect{M}\cdot\hat{x}\neq0$} & \multirow{2}{5cm}{$Z\neq0,\vect{M}\cdot\hat{y}\neq0$ and $\vect{M}\cdot\hat{z}\neq0$} \\
 & $d_{xy}$ & $k_{x}\hat{x}\pm k_{y}\hat{y}$ &  & \\
\hline\hline
\end{tabular}
\end{center}
\end{table}

\begin{figure}[h]
\begin{center}
\onefigure[height=4cm, width=8cm]{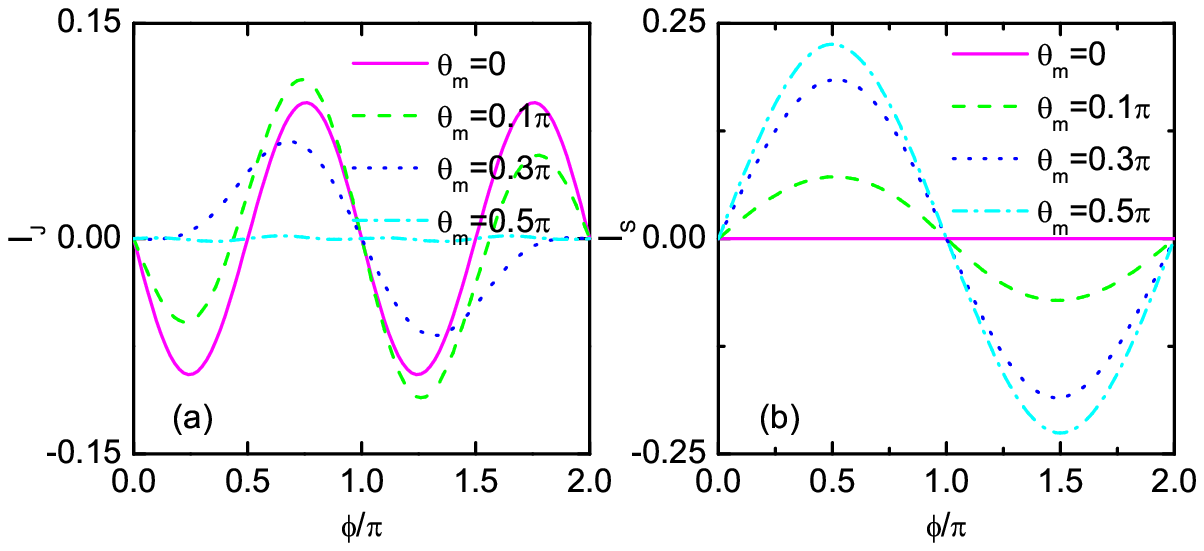}
\caption{(Colour online) (a): The CCPRs for $\phi_{m}=0$, $Z=1$, $X=1$ and $T=0.3T_{C}$ in the $d_{xy}$-wave SS/ HP$_{1+}$-wave TS junction. (b): The corresponding SCPRs with the same parameters.}\label{fig.5}
\end{center}
\end{figure}

We turn to the junctions with the finite non-magnetic potential, i.e., $Z\neq0$. We take the $d_{xy}$-wave SS/FI/ HP$_{1+}$-wave TS junction as an example. Fig.~\ref{fig.5} shows the CCPRs and the SCPRs with $\phi_{m}=0$ for various $\theta_{m}$ at $T=0.3T_{C}$. In accordance with the rules in table~\ref{table.1}, there will be no $\cos{\phi}$-type LOC both in the CCPRs and in the SCPRs. However, when the magnetization possesses both the $x$ and $z$-component, such as $\theta_{m}=0.1\pi$, $I_{J}\vert_{\phi=0.5\pi}>0$ and $I_{J}\vert_{\phi=1.5\pi}<0$ indicate the existence of the $\sin{\phi}$-type LOC in the charge current. Especially when $\theta_{m}=0.3\pi$, the sinusoidal curve starts to dominate the CCPR. Actually, $j(w_{n},\theta)$ is proportional to the combination of $\sin{\phi}$ and $\sin{2\phi}$ when $0<\theta_{m}<0.5\pi$. Fig. 6 gives the CCPRs and the SCPRs with $\phi_{m}=0.5\pi$. The magnetization has no $x$-component, i.e. $\vect{M}\cdot\hat{x}=0$. The emergence of the $\cos{\phi}$-type LOC in the charge current at $\theta_{m}\neq0$ and in the spin current at $\theta_{m}\neq0$ and $0.5\pi$ fits the rules in table~\ref{table.1}. $I_{J}\vert_{\phi=0.5\pi}=0$ and $I_{J}\vert_{\phi=1.5\pi}=0$ demonstrate the absence of the $\sin{\phi}$-type LOC. Fig. 7 presents the CCPRs and the SCPRs with $\theta_{m}=0.5\pi$ for various $\phi_{m}$. The magnetization in this case is in the $xy$-plane and has no $z$-component, i.e. $\vect{M}\cdot\hat{z}=0$. There is no $\sin{\phi}(\cos{\phi})$-type current in the CCPRs (SCPRs). When the magnetization sweep across the $x$-axis, the charge current with the $\cos{\phi}$ form will reverse its sign.

From the above discussions, we can derive the selection rules for the $\sin{\phi}$-type LOC in the charge current and the $\cos{\phi}$-type LOC in the spin current which have been summarised in table~\ref{table.1}. The non-magnetic potential with the non-zero value is an indispensable condition. Actually, the non-magnetic potential not only leads to the presence of the LOC with the sinusoidal form in the charge current but also modifies the symmetries of CCPRs. The equalities in eq.~(\ref{eq10}), (\ref{eq14}) and (\ref{eq16}) do not hold when $Z\neq0$.

The selection rules in table~\ref{table.1} reveal the interaction forms of the helical superconductivity and ferromagnetism which are involved in the free energy function $F$. We consider the product of the order parameter in SS and the $\vect{d}$-vector in TS which can be written as
\begin{equation}
g_{1}(k_{x},k_{y})\hat{x}+g_{2}(k_{x},k_{y})\hat{y}.
\end{equation}
The functions $g_{1}(k_{x},k_{y})$ and $g_{2}(k_{x},k_{y})$ have opposite parity for $k_{x}$ and $k_{y}$. The selection rules indicate that $M_{x}\sin{\phi}$ and $M_{y}M_{z}\cos{\phi}$ will contribute to the free energy if $g_{1}(k_{x},k_{y})$ is odd for $k_{x}$ and $g_{2}(k_{x},k_{y})$ is odd for $k_{y}$; $M_{y}\sin{\phi}$ and $M_{x}M_{z}\cos{\phi}$ will contribute to the free energy if $g_{1}(k_{x},k_{y})$ is odd for $k_{y}$ and $g_{2}(k_{x},k_{y})$ is odd for $k_{x}$. The non-zero conditions for the terms in $\frac{\partial F}{\partial \phi}$ are just the LOC selection rules for the charge current; the non-zero conditions for the terms in the $z$ component of $\vect{M}\times\frac{\partial F}{\partial \vect{M}}$ \cite{Brydon6} are just the the LOC selection rules for the spin current. The symmetries and sign reversal satisfied by the spin current can be derived from the expression of the free energy. The above derivation of the selection rules is universal for the SS/FI/TS junctions with helical states which also applies to the states including more components such as $(k_{x}^2-k_{y}^2)(k_{x}\hat{x}\pm k_{y}\hat{y})$, $(k_{x}^2-k_{y}^2)(k_{y}\hat{x}\pm k_{x}\hat{y})$, $k_{x}k_{y}(k_{x}\hat{x}\pm k_{y}\hat{y})$ and $k_{x}k_{y}(k_{y}\hat{x}\pm k_{x}\hat{y})$ \cite{Blount, Yip}.
\section{Conclusion}
The in-plane $\vect{d}$-vector in the superconductor with helical states and its wavevector-dependent direction can bring novel and rich physics due to the interplay between superconductivity and ferromagnetism, which can be displayed in the electronic transport properties and the Josephson effect. In this paper, we study the CCPRs and the SCPRs in SS/FI/TS Josephson junctions. The gap function in SS is assumed to be $s$, $d_{x^2-y^2}$ or $d_{xy}$-wave symmetry. Both $\cos{\phi}$-type and $\sin{\phi}$-type LOC exist in the junctions as long as the selection rules are satisfied. The rules themselves have close connection to the gap functions. The sign reversal and symmetries of the CCPRs and the SCPRs are revealed.

\acknowledgments
This work is supported in part by Qingdao Science and Technology Program (Grant No. 14-2-4-110-JCH) and by the National Natural Science Foundation of China (Grant Nos. 11447175 and 11547035).

\section{Figure captions}

Fig. 1: (Colour online) (a): Schematic illustration of SS/FI/TS junctions. The $x (y)$-axis is defined by the crystallographic $a (b)$-axis. (b): The magnetization in FI specified by the polar angle $\theta_{m}$ and the azimuthal angle $\phi_{m}$. (c): The spins of Cooper pairs for the helical states which can be thought of as the superposition of the two states with spin parallel and anti-parallel to the $z (c)$-axis.

Fig. 2: (Colour online) The CCPRs for $\phi_{m}=0$, $Z=0$, $X=1$ and $T=0.3T_{C}$ in the $s$-wave SS/HP$_{1\pm}$-wave TS junctions.

Fig. 3: (Colour online) (a): The CCPRs for $\phi_{m}=0.5\pi$, $Z=0$, $X=1$ and $T=0.3T_{C}$ in the $s$-wave SS/ HP$_{1\pm}$-wave TS junctions. (b) The corresponding SCPRs with the same parameters.

Fig. 4: (Colour online) (a): The CCPRs for $\theta_{m}=0.5\pi$, $Z=0$, $X=1$ and $T=0.3T_{C}$ in the $s$-wave SS /HP$_{1\pm}$-wave TS junctions. (b): The corresponding SCPRs with the same parameters.

Fig. 5: (Colour online) (a): The CCPRs for $\phi_{m}=0$, $Z=1$, $X=1$ and $T=0.3T_{C}$ in the $d_{xy}$-wave SS/ HP$_{1+}$-wave TS junction. (b): The corresponding SCPRs with the same parameters.

Fig. 6: (Colour online) (a): The CCPRs for $\phi_{m}=0.5\pi$, $Z=1$, $X=1$ and $T=0.3T_{C}$ in the $d_{xy}$-wave SS/ HP$_{1+}$-wave TS junction. (b): The corresponding SCPRs with the same parameters.

Fig. 7: (Colour online) (a): The CCPRs for $\theta_{m}=0.5\pi$, $Z=1$, $X=1$ and $T=0.3T_{C}$ in the $d_{xy}$-wave SS/ HP$_{1+}$-wave TS junction. (b): The SCPRs with the same $\theta_{m}$, $Z$, $X$ and $T$.

\begin{figure}[h]
\begin{center}
\onefigure[height=4cm, width=8cm]{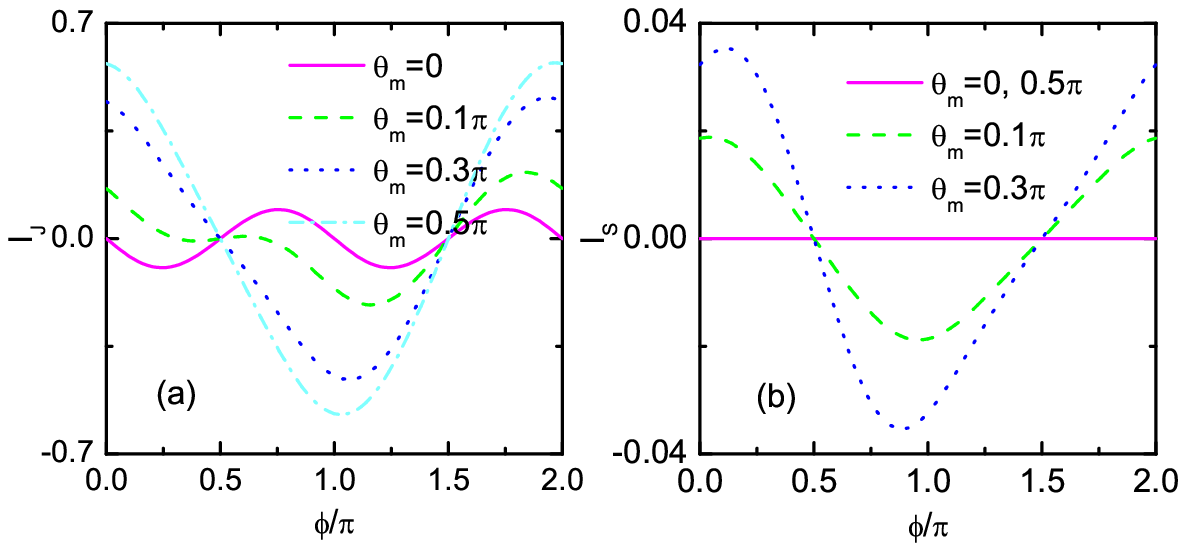}
\caption{(Colour online) (a): The CCPRs for $\phi_{m}=0.5\pi$, $Z=1$, $X=1$ and $T=0.3T_{C}$ in the $d_{xy}$-wave SS/ HP$_{1+}$-wave TS junction. (b): The corresponding SCPRs with the same parameters.}\label{fig.6}
\end{center}
\end{figure}

\begin{figure}[h]
\begin{center}
\onefigure[height=4cm, width=8cm]{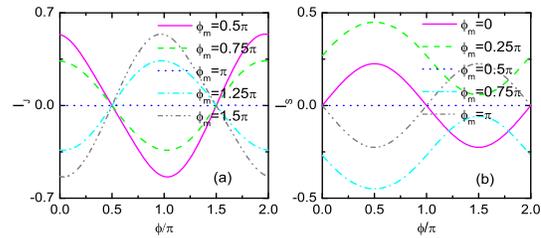}
\caption{(Colour online) (a): The CCPRs for $\theta_{m}=0.5\pi$, $Z=1$, $X=1$ and $T=0.3T_{C}$ in the $d_{xy}$-wave SS/ HP$_{1+}$-wave TS junction. (b): The SCPRs with the same $\theta_{m}$, $Z$, $X$ and $T$.}\label{fig.7}
\end{center}
\end{figure}

\end{document}